\documentclass[preprint,showpacs,showkeys,preprintnumbers,amsmath,amssymb]{revtex4}

\usepackage{graphicx}
\usepackage{dcolumn}
\usepackage{bm}

\begin{document}

\preprint{APS/123-QED}

\title{Role of Noise in a Market Model with Stochastic Volatility}

\author{Giovanni Bonanno, Davide Valenti$^\circ$ and Bernardo Spagnolo$^\dagger$}
 \affiliation{Dipartimento di Fisica e Tecnologie Relative, Group
of Interdisciplinary Physics \footnote{Electronic address:
http://gip.dft.unipa.it},
 Universit\`a di Palermo\\ Viale delle Scienze, I-90128 Palermo,
Italy \\ $^\circ$ valentid@gip.dft.unipa.it, $^\dagger$
spagnolo@unipa.it}
\date{\today}

\begin{abstract}
We study a generalization of the Heston model, which consists of
two coupled stochastic differential equations, one for the stock
price and the other one for the volatility. We consider a cubic
nonlinearity in the first equation and a correlation between the
two Wiener processes, which model the two white noise sources.
This model can be useful to describe the market dynamics
characterized by different regimes corresponding to normal and
extreme days. We analyze the effect of the noise on the
statistical properties of the escape time with reference to the
noise enhanced stability (NES) phenomenon, that is the noise
induced enhancement of the lifetime of a metastable state. We
observe NES effect in our model with stochastic volatility. We
investigate the role of the correlation between the two noise
sources on the NES effect.
\end{abstract}

\maketitle

\section{\label{sec:intro}Introduction}
The presence of noise in physical systems is a well known
phenomenon. It is common opinion that noise affecting the dynamics
of a system introduces some degree of instability to the system
itself, but there is evidence that in some cases the noise can
increase the stability of a
system~\cite{NESChaos-TD,NESPiecewise,NESothers,NESreview}. Noise
modeling through the use of the stochastic processes formalism has
applications that involve many systems including physics, biology,
ecology~\cite{GenericNoise,MBE,Ecology} and even financial
markets~\cite{Mand,MSt,Bou}. The most basic model for financial
market is the geometric Brownian motion~\cite{Hull}. This model has
different drawbacks, it cannot reproduce in fact three important
stylized facts observed in financial time series: (i) the non
Gaussian distribution of returns, (ii) the fat tails~\cite{MSt,Bou},
and (iii) the stochastic character of volatility, which is
characterized by long range memory and
clustering~\cite{MSt,Bou,Bouchaud,Dacorogna}. More complex models
have been developed to reproduce the dynamics of the volatility. It
is worthwhile citing the ARCH~\cite{Arch} and GARCH~\cite{Garch}
models, where the actual volatility depends on the past values of
squared return (ARCH) and also on the past values of the volatility
(GARCH). Another class of models use a system of stochastic
equations writing the price as a geometric Brownian motion coupled
with a non-constant volatility described by a second stochastic
differential equation. The Heston model uses for the volatility a
multiplicative stochastic process characterized by mean
reversion~\cite{Heston,Hull-White}. Both the models presented so far
have exponential autocorrelation function so they are not able to
reproduce quantitatively the long range memory observed in real
markets. Nonetheless using values of the characteristic time that
are sufficiently high, they are able to give accurate statistic for
the stock prices, by tuning only few parameters. Another important
characteristic of financial markets is the presence of different
regimes. Markets indeed present days of normal activity and extreme
days where very high or very low price variations can be observed.
These are known as crash and rally days. A nonlinear Langevin market
model has already been proposed~\cite{BouchaudCont}, where different
regimes are modelled by means of an effective potential for price
returns. In some circumstances this potential has a cubic shape with
a metastable state and a potential barrier. The dynamics inside the
metastable state represents the days of normal evolution while the
escape after the potential barrier represents the beginning of a
crisis. Metastable states are ubiquitous in physics and the effect
of noise in such systems has been extensively
studied~\cite{NESChaos-TD,NESPiecewise,NESothers,NESreview}, but
considering the noise intensity as a parameter (physical models
described by additive stochastic differential equations for
example). Financial markets with their stochastic volatility are an
example of systems where the noise intensity is far from being a
constant parameter, but it is indeed a stochastic process itself.
Moreover there is evidence in nature that the noise intensity is not
a constant parameter and can be modelled as a multiplicative
noise~\cite{MBE,Ecology,OltreNES}. So it is interesting to release
the hypothesis of parametric noise intensity in financial market
models as well as in natural systems.

\section{\label{sec:nes_heston}The Heston model with a metastable state}

The Heston model introduced in the previous section is described
by the following system of coupled stochastic differential
equations~\cite{Heston}

\begin{eqnarray}
  dx(t) & = & (\mu - v(t)/2) \cdot dt + \sqrt{v(t)} \cdot dZ(t) \nonumber \\
  dv(t) & = & a(b-v(t)) \cdot dt + c \sqrt{v(t)} \cdot dW(t)
\label{Eqn:Heston}
\end{eqnarray}
The price $p(t)$ follows a geometric random walk whose standard
deviation is another stochastic process. Here $x(t)= \ln p(t)$ is
the $log$ of the price, $Z(t)$ and $W(t)$ are uncorrelated Wiener
processes with the usual statistical properties: (i) $\langle
dZ(t) \rangle =0$ and $\langle dZ(t) \cdot dZ(t') \rangle =
\delta(t-t') dt$; (ii) $\langle dW(t) \rangle =0$ and $\langle
dW(t) \cdot dW(t') \rangle = \delta(t-t') dt$. The $v(t)$ process
is characterized by mean reversion, i.e. its deterministic
solution has an exponential transient with characteristic time
equal to $a^{-1}$, after which the process tends to its asymptotic
value $b$. The process for $v(t)$ exhibits the phenomenon of
volatility clustering, alternating calm with burst periods of
volatility, and has an exponential autocorrelation function. The
smaller the value of $a$ the longer are the bursts in volatility.
Heston model has been subject of recent investigation by
econophysicists~\cite{BonannoPhysicaA,Yakovenko,BonannoFNL,Silva}.
 The equations of the system $(1)$ are well known in finance, they
 represent respectively the log-normal geometric Brownian motion
 stock process used by Black and Scholes for option pricing~\cite{BS,
 Merton}, and the Cox-Ingersoll-Ross (CIR) mean-reverting stochastic
 differential equation first introduced for interest rate
 models~\cite{Cox,Chalasani}.

\begin{figure}[htbp]
\vspace{5mm}
\centering{\resizebox{9cm}{!}{\includegraphics{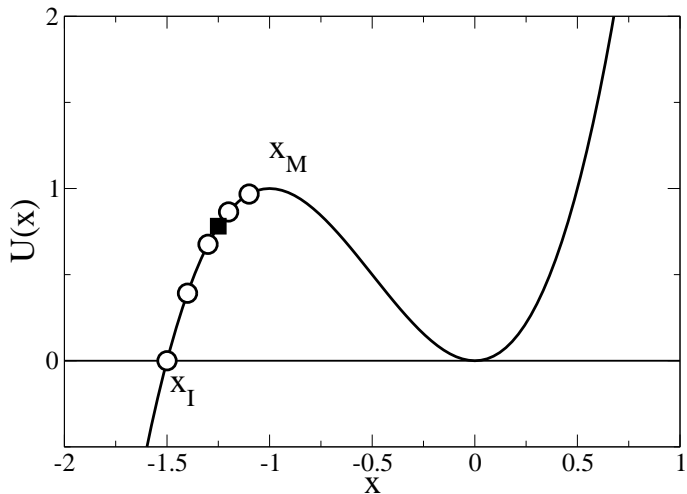}}}
\caption{\label{Fig:Cubico} Cubic potential used in the dynamical
equation for the price x(t). The points in the figure indicate the
starting positions used in our simulations.}
\end{figure}

Here we consider a generalization of the Heston model, by replacing
the geometric Brownian motion with a random walk in the presence of
a cubic nonlinearity. This generalization represents a
"\emph{Brownian particle}" moving in an \emph{effective} potential
with a metastable state, in order to model those systems with two
different dynamical regimes like financial markets in normal
activity and extreme days~\cite{BouchaudCont}. The equations of the
new model are

\begin{eqnarray}
  dx(t) & = & - \left(\frac{\partial U}{\partial x} +
  \frac{v(t)}{2}\right) \cdot dt + \sqrt{v(t)}
  \cdot dZ(t) \nonumber \\
  dv(t) & = & a(b-v(t)) \cdot dt + c \sqrt{v(t)} \cdot dW(t),
\label{Eqn:BS}
\end{eqnarray}
where $U$ is the \emph{effective} cubic potential $U(x)=px^3+qx^2$,
with $p=2$ and $q=3$ (see Fig.~\ref{Fig:Cubico}), $Z(t)$ and $W(t)$
are standard Wiener processes.
Let us call $x_M$ the abscissa of the potential maximum and $x_I$
the abscissa where the potential intersects the $x$ axes. The
intervals $x_o < x_I$ and $I=[x_I,x_M]$ are clearly regions of
instability for the system. In systems with a metastable state like
this, the noise can originate interesting effects that increase
instead of decrease the stability, by enhancing the lifetime of the
metastable
state~\cite{NESChaos-TD,NESPiecewise,NESothers,NESreview}. A good
example is the Noise Enhanced Stability (NES) phenomenon. The mean
escape time $\tau$ for a Brownian particle moving throughout a
barrier $\Delta U$, with a noise intensity $v$, is given by the well
known exponential Kramers law~\cite{Hanggi,Gardiner}

\begin{equation}
  \tau =  A exp\left[{\frac{\Delta U}{v}}\right],
\label{Eqn:Kramers}
\end{equation}
where $\tau$ is a monotonically decreasing function of the noise
intensity $v$, and $A$ is a prefactor which depends on the potential
profile. This is true only if the random walk starts from initial
positions inside the potential well. When the starting position is
chosen in the instability region $x_o < x_M$, $\tau$ exhibits an
enhancement behavior, with respect to the deterministic escape time,
as a function of $v$. Particularly for initial positions $x_0 <
x_I$, we have nonmonotonic behavior of $\tau$ as a function of
$v$~\cite{NESPiecewise,NES03}. This is the NES effect and can be
explained considering the barrier "\emph{seen}" by the Brownian
particle starting at the initial position $x_0 $, that is $\Delta
U_{in} = U(x_{max})-U(x_0)$. Moreover $\Delta U_{in}$ is less than
$\Delta U$ as long as the starting position $x_0$ lyes into the
interval $I=[x_I,x_M]$. Therefore for a Brownian particle, from a
probabilistic point of view, it is easier to enter into the well
than to escape from, when the particle is entered. So a small amount
of noise can increase the lifetime of the metastable
state~\cite{NESChaos-TD,NESPiecewise,NESreview,NES03}. For a
detailed discussion on this point and different dynamical regimes
see Refs.~\cite{NESPiecewise,NES03}. When $v$ is much greater than
$\Delta U$, the Kramers behavior is recovered. The NES effect has
been experimentally observed in a tunnel diode and theoretically
predicted in a wide variety of systems such as for example chaotic
map, Josephson junctions, chemical reaction kinetics, and neuronal
dynamics models~\cite{NESChaos-TD,NESPiecewise,NESothers,NESreview}.
Our modified Heston model, characterized by a stochastic volatility
and a nonlinear Langevin equation for the returns, has two limit
regimes, corresponding to the cases $a=0$, with only the noise term
in the equation for the volatility $v(t)$, and $c=0$ with only the
reverting term in the same equation. This last case corresponds to
the usual parametric constant volatility case. In fact, apart from
an exponential transient, the volatility reaches the asymptotic
value $b$. The NES effect should be observable in the latter case as
a function of $b$, which is the average volatility. In this case, in
fact, we have the motion of a Brownian particle in a fixed cubic
potential with a metastable state and an enhancement of its lifetime
for particular initial conditions (see
Refs.~\cite{NESPiecewise,NES03}).

\section{\label{sec:correlation}Enhancement of the escape time}

The two processes of Eqs.~(\ref{Eqn:Heston}) and (\ref{Eqn:BS}) are
actually uncorrelated. In financial markets the two processes can be
correlated, and a negative correlation between the processes is
known as \emph{leverage effect}~\cite{Fouque}. Heston model with
correlation has been recently discussed in the scientific
literature~\cite{Yakovenko,Silva}. Our modified Heston model becomes
therefore

\begin{eqnarray}
  dx(t)   & = & - \left(\frac{\partial U}{\partial x} +
  \frac{v(t)}{2}\right) \cdot dt + \sqrt{v(t)} \cdot dZ(t) \nonumber \\
  dv(t)   & = & a(b-v(t)) \cdot dt + c \sqrt{v(t)} \cdot dW_c(t) \nonumber \\
  dW_c(t) & = & \rho \cdot dZ(t) + \sqrt{1-\rho^2} \cdot dW(t),
\label{Eqn:BSCorr}
\end{eqnarray}
where $Z(t)$ and $W(t)$ are uncorrelated Wiener processes as in
Eqs.~(\ref{Eqn:Heston})~and~(\ref{Eqn:BS}), and $\rho$ is the cross
correlation coefficient between the noise sources. The investigation
is performed simulating the process of Eqs.~(\ref{Eqn:BSCorr}) with
time integration step $\Delta t=0.01$, and for the fixed starting
position $x_0=-1.25$ in the $I$ interval (this initial position is
shown as a black square point in Fig.~\ref{Fig:Cubico}). The
absorbing barrier is located at $x=-6.0$, and the results are
averaged over $10^5$ escape events. The algorithm used to simulate
the noise sources in Eqs.~(\ref{Eqn:BSCorr}) is based on the
\emph{Box-Muller} method for generating random processes with a
Gaussian distribution. The numerical integration of
Eqs.~(\ref{Eqn:BSCorr}) is done by using the \emph{forward Euler}
method~\cite{NumRec}.

\begin{figure}[htbp]
\vspace{5mm}
\centering{\resizebox{9cm}{!}{\includegraphics{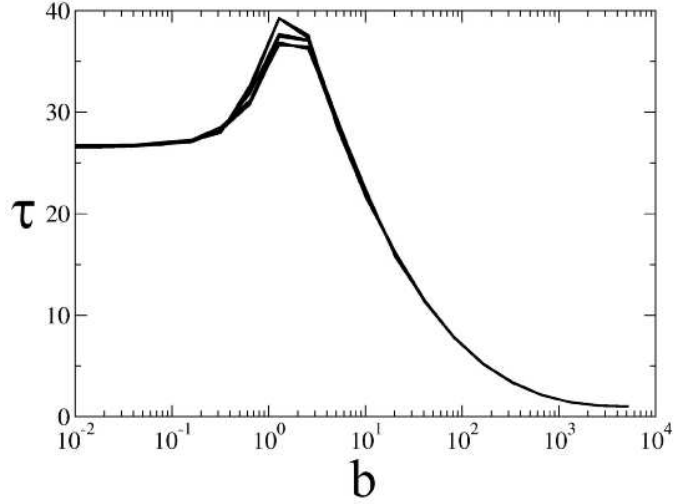}}}
\caption{\label{Fig:t_b_corr} Mean escape time $\tau$ as a function
of the mean volatility $b$. The other parameters have the fixed
values $a=10^{-2}$ and $c=10^{-2}$. The six curves correspond to the
following  values of the correlation: $\rho=-0.8, -0.5, -0.1, 0.1,
0.5, 0.8$. The fixed starting position is $x_0 = -1.25$. The values
of the potential parameters (see Eq.~(\ref{Eqn:BS})) are: $p = 2$,
$q = 3$. }
\end{figure}

Our first result shows that the curve $\tau$~vs.~$b$ is weakly
dependent on the value of the $\rho$ parameter. This is shown in
Fig.~\ref{Fig:t_b_corr}, where all the curves correspond to the
region of the parameters space where the effect is observable. There
is indeed a weak variation in the maximum value. The highest maximum
values correspond to the highest absolute $\rho$ values. It is worth
noting that the correlation affects directly the noise term in the
$v(t)$ equation, and it has negligible influence on the reverting
term of the same equation.

\begin{figure}[htbp]
\vspace{13mm}
\centering{\resizebox{12cm}{!}{\includegraphics{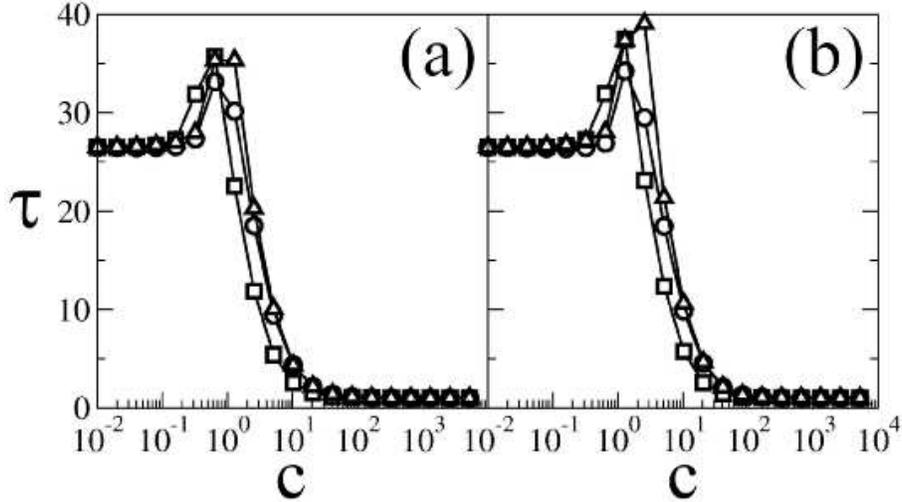}}}
\caption{\label{Fig:t_c_corr} Mean escape time $\tau$ as a function
of the model noise intensity $c$ for two different values of the
parameter $a$: (a) $a=2.0$ and (b) $a=20$. The value of the
parameter $b$ is fixed to $b=10^{-2}$. The fixed starting position
$x_0$ and the potential parameters $p$ and $q$ are the same of
Fig.~\ref{Fig:t_b_corr}. The different curves correspond to the
following values of $\rho$: -0.8 (circle), 0.0 (square), 0.8
(triangle).}
\end{figure}

\begin{figure}[htbp]
\vspace{5mm}
\centering{\resizebox{12cm}{!}{\includegraphics{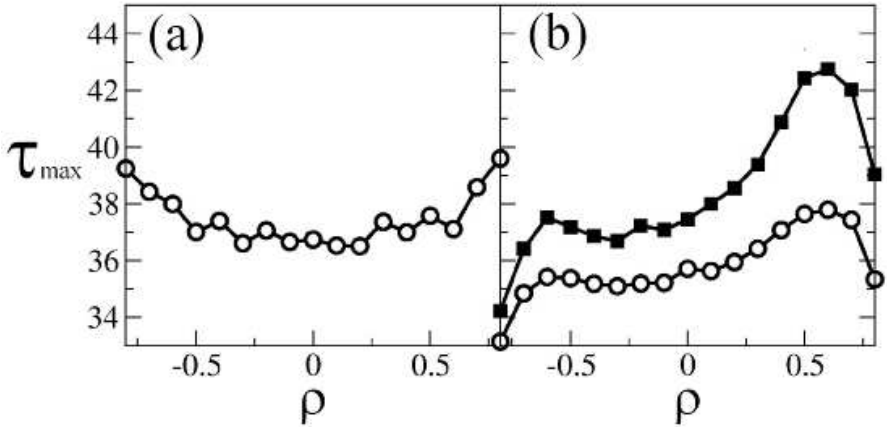}}}
\caption{\label{Fig:max_rho} The values of $\tau_{max}$ as a
function of the correlation coefficient of the curves reported in
Figs.~\ref{Fig:t_b_corr} and~\ref{Fig:t_c_corr}. Specifically: (a)
$\tau_{max}$ from Fig.~\ref{Fig:t_b_corr} ($\tau$ as a function of
parameter b) with $a = 10^{-2}$, $c = 10^{-2}$); (b) $\tau_{max}$
from Fig.~\ref{Fig:t_c_corr} with $a = 2.0$~(circles) and $a =
20$~(squares), $b = 10^{-2}$. The fixed starting position $x_0$ and
the potential parameters $p$ and $q$ are the same of
Fig.~\ref{Fig:t_b_corr}.}
\end{figure}

The curves of $\tau~vs.~c$ however have an evident dependence on the
parameter $c$. This is shown in Fig.~\ref{Fig:t_c_corr}, for two
values of parameter $a$ for which the NES effect is observable. The
correlation affects the position as well as the value of the maximum
of $\tau$ (see Fig.~\ref{Fig:t_c_corr}), but in a different way. The
position of the maximum increases only for very high positive values
of the correlation coefficient. The maximum value of $\tau$
increases with positive correlation but decreases with negative
correlation. This effect is more evident for the higher value of
$a$, as shown in Fig.~\ref{Fig:t_c_corr}b.
To comment this result we first note that a negative correlation
between the logarithm of the price and the volatility means that a
decrease in $x(t)$ induces an increase in the volatility $v(t)$, and
this causes the Brownian particle to escape easily from the well. As
a consequence the mean lifetime of the metastable state decreases,
even if the nonmonotonic behavior is still observable. On the
contrary, when the correlation $\rho$ is positive, the Brownian
particle stays more inside the well, decrease in $x(t)$ indeed is
associated with decrease in the volatility. The escape process
becomes slow and this increases further the lifetime of the
metastable state, causing an increase in the value of the maximum of
the curve of Fig.~\ref{Fig:t_c_corr}. To illustrate better this
aspect and the behaviour of Figs.~\ref{Fig:t_b_corr}
and~\ref{Fig:t_c_corr} near the maximum, we plot the values
$\tau_{max}$ of the maximum as a function of the correlation
coefficient $\rho$ and we show these curves in
Fig.~\ref{Fig:max_rho}. Specifically in Fig.~\ref{Fig:t_b_corr}a we
report the values of $\tau_{max}$ related to the curves of $\tau$ of
Fig.~\ref{Fig:t_b_corr}, and in Fig.~\ref{Fig:t_b_corr}b those
related to Fig.~\ref{Fig:t_c_corr}.
The increase of stability is evident in panel (b) as $\tau_{max}$
increases with $\rho$.

\begin{figure}[htbp]
\vspace{5mm}
\centering{\resizebox{9cm}{!}{\includegraphics{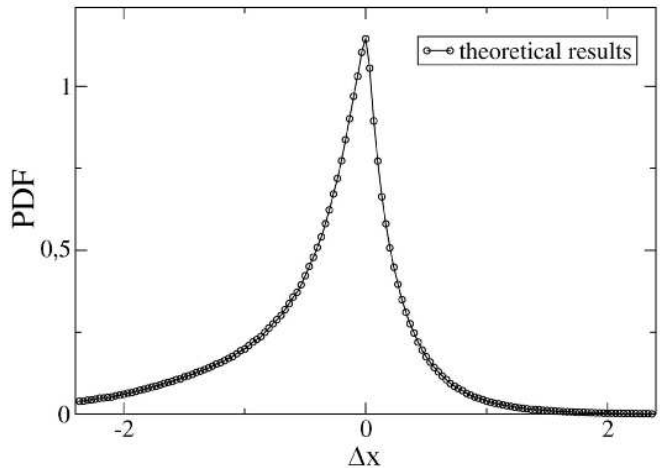}}}
\caption{\label{Fig:PDF-returns} Probability density function of
stock price returns for the Heston model with a metastable state and
correlated noise sources (Eqs.~(\ref{Eqn:BSCorr})). The parameters
of the Cox-Ingersoll-Ross process of Eqs.~(\ref{Eqn:BSCorr}) are: $a
=2$, $b = 0.01$, $c = 26$. The cross correlation coefficient is
$\rho = -0.3$. The potential parameters are the same of
Fig.~\ref{Fig:t_b_corr}.}
\end{figure}

It is interesting to show the probability density function (PDF) of
stock price returns for the model described by the
Eqs.~(\ref{Eqn:BSCorr}). This is done in Fig.~\ref{Fig:PDF-returns}.
As one can see the qualitative behavior of a fat tail distribution,
typical of real financial market data, is recovered~\cite{MSt,Bou},
but with a peculiar asymmetry. To characterize quantitatively the
PDF of returns (Fig.~\ref{Fig:PDF-returns}) as regards the width,
the asymmetry and the fatness of the distribution, we calculate the
mean value $<\Delta x>$, the variance $\sigma_{\Delta x}$, the
skewness $\kappa_3$ and the kurtosis $\kappa_4$. We obtained the
following values: $<\Delta x> = - 0.434$, $\sigma_{\Delta x} =
0.903$, $\kappa_3 = - 2.086$, $\kappa_4 = 9.417$. These statistical
quantities clearly show the asymmetry of the distribution and its
leptokurtic nature observed in empirical investigations,
characterized by a narrow and larger maximum, and by fatter tails
than in the Gaussian distribution~\cite{MSt,Bou}.

The presence of the asymmetry is very interesting and it will be
subject of future investigations. It is worthwhile to note, however,
that the PDF of returns become asymmetric in crash and rally
days~\cite{Lil00,LiMa00}, that are just the time periods of
financial data where the related dynamical regimes could be
described by the models~(\ref{Eqn:BS}) and~(\ref{Eqn:BSCorr}). Of
course the quantitative agreement between the PDF of real data and
that obtained from these models requires further investigations on
the parameter value choice of the cubic potential, the parameters of
the simple Heston model and the correlation coefficient. This
analysis is outside the aim of the present work and will be further
investigated in a forthcoming paper.

Finally in the following Fig.~\ref{Fig:PDF-escape times} we report
the comparison between the probability density function of the
escape times of daily price returns from real market data and that
obtained from the model described by the Eqs.~(\ref{Eqn:BSCorr}).
The data set used here consists of daily closure prices for 1071
stocks traded at the NYSE and continuously present in the 12-year
period $1987-1998$ (3030 trading days). The same data were used in
previous investigations by one of the
authors~\cite{BonannoPhysicaA,Bon_PRE_03,Bon_EPJB04}. From this data
set we obtained the time series of the returns and we calculated the
time to hit a fixed threshold starting from a fixed initial
position. The two thresholds were chosen as a fraction of the
standard deviation $\sigma_n$ observed for each stock during the
above mentioned whole time period (n is the stock index, varying
between $1$ and $1071$). Specifically we chose: $(\Delta x_i)_n =
-0.1 \sigma_n$ and  $(\Delta x_f)_n = -1.0 \sigma_n$. The parameters
of the CIR process are: $a = 20$, $b = 0.01$, $c = 2.4$. The cross
correlation coefficient is $\rho = -0.9$, and the potential
parameters are the same of Fig.~\ref{Fig:t_b_corr}. As one can see
the agreement between real and theoretical data is very good, except
at small escape times. The choice of this parameter data set is not
based on a fitting procedure as that used for example in
Ref.~\cite{Yakovenko}, where the minimization of the mean square
deviation between the PDF of the returns extracted from financial
data and that obtained theoretically is done. We chose the parameter
set in the range in which we observe the nonmonotonic behaviour of
the mean escape time as a function of the parameters $b$ and $c$.
Then by a trial and error procedure we selected the values of the
parameters $a$, $b$, and $c$ for which we obtain the best fitting
between the PDF of escape times of the price returns calculated from
the modified Heston model (Eqs.~(\ref{Eqn:BSCorr})) and that
obtained from time series of real market data. Of course a better
quantitative procedure could be done, by considering also the
potential parameters. This will be done, together with a detailed
analysis of PDF of returns and its asymmetry, in a forthcoming
paper.

\begin{figure}[htbp]
\vspace{5mm}
\centering{\resizebox{9cm}{!}{\includegraphics{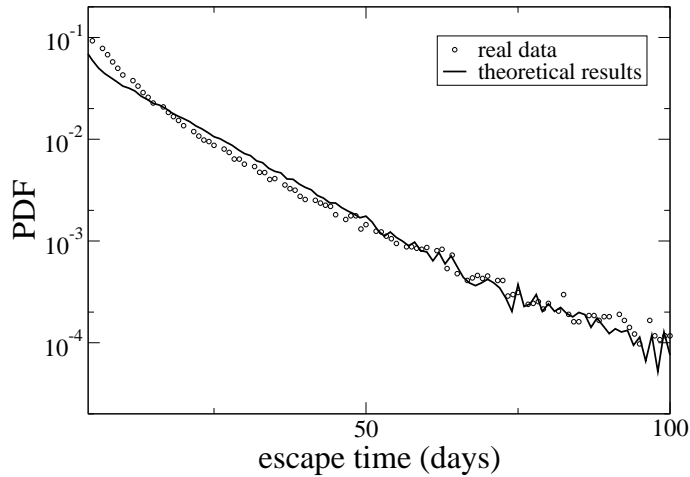}}}
\caption{\label{Fig:PDF-escape times} Comparison between the
probability density function of escape times of the price returns
for the modified Heston model with a metastable state
(Eqs.~(\ref{Eqn:BSCorr})) (solid line) and the PDF of escape times
of returns for real market data (cirles). The parameter values of
the model~(\ref{Eqn:BSCorr}) are: $p = 2$, $q = 3$, $a = 20$, $b =
10^{-2}$, and $c = 2.4$. The cross correlation coefficient between
the noise sources is: $\rho = -0.9$.}
\end{figure}

\section{Conclusions}

We have investigated the statistical properties of the escape time
in a generalized Heston market model, characterized by the presence
of a metastable state in the effective potential of the logarithm of
the price $p(t)$. We observe the NES effect in the system
investigated. The presence of correlation between the stochastic
volatility and the noise source which affects directly the dynamics
of the quantity $x(t) = \ln p(t)$ (as in usual market models), can
influence the stability of the market. Specifically a positive
correlation between $x(t)$ and volatility $v(t)$ slows down the
walker escape process, that is it delays the crash phenomenon by
increasing the stability of the market. A negative correlation on
the contrary accelerates the escape process, lowering the stability
of the system.
\\
\\
\\
\noindent This work was partially supported by MIUR.


\begin{thebibliography}{20}

\bibitem{NESChaos-TD}
 R.N. Mantegna and B. Spagnolo, Phys. Rev. Lett.
{\bf 76}, 563, (1996); J.E. Hirsch, B.A. Huberman and D.J.
Scalapino Phys. Rev. A {\bf 25}, 519, (1982).

\bibitem{NESPiecewise}
N.V. Agudov and B. Spagnolo, Phys. Rev. E {\bf 64}, 035102
(\textbf{R}), (2001).

\bibitem{NESothers}
A. N. Malakhov, A. L. Pankratov, Physica C \textbf{269}, 46
(1996); D. Dan, M. C. Mahato, A. M. Jayannavar, Phys. Rev. E
\textbf{60}, 6421 (1999); M. Yoshimoto, Phys. Lett. A
\textbf{312}, 59 (2003).

\bibitem{NESreview}
A.L. Pankratov and B. Spagnolo, Phys. Rev. Lett. \textbf{93},
177001 (2004); Bernardo Spagnolo, Alexander A. Dubkov, Nikolai V.
Agudov, Acta Physica Polonica B, Vol. \textbf{35} (4), 1419; A. A.
Dubkov, N. V. Agudov and B. Spagnolo, Phys. Rev. E \textbf{69},
061103 (2004); Evgeniya V. Pankratova, Andrey V. Polovinkin, and
Bernardo Spagnolo, Physics Letters A, 344 (1), 43-50 (2005).

\bibitem{GenericNoise}
Peter H$\ddot{a}$nggi and Fabio Marchesoni, {\em Chaos}
\textbf{15}, 026101/1-5 (2005).

\bibitem{MBE} B. Spagnolo, D. Valenti, A. Fiasconaro, Math.
Biosciences and Engineering \textbf{1}, 185 (2004).

\bibitem{Ecology} See the special section on \emph{"Complex Systems"}, Science
\textbf{ 284}, 79-107 (1999); O. N. Bjornstad and B. T. Grenfell,
Science \textbf{293}, 638 (2001).

\bibitem{Mand}
B. B. Mandelbrot, {\em Fractals and Scaling in Finance}, Springer,
New York (1997).

\bibitem{MSt}
R.N. Mantegna and H.E. Stanley {\em An introduction to econophysics:
correlations and complexity in finance}, (Cambridge University
Press, Cambridge, 2000).

\bibitem{Bou}
J.P. Bouchaud and M. Potters {\em Theory of financial risks},
(Cambridge University Press, Cambridge, 2000).

\bibitem{Bouchaud}
Lisa Borland, Jean-Philippe Bouchaud, Jean-Francois Muzy, Gilles
Zumbach, {\em The Dynamics of Financial Markets -- Mandelbrot's
multifractal cascades, and beyond}, cond-mat/0501292 (2005); Lisa
Borland and Jean-Philippe Bouchaud, {\em On a multi-timescale
statistical feedback model for volatility fluctuations},
physics/0507073 (2005).

\bibitem{Hull}
J.~C. Hull, {\em Options, Futures, and Other Derivatives},
Prentice-Hall, New Jersey, (1997).

\bibitem{Dacorogna}
M.M. Dacorogna, R. Gencay, U.A. M\"uller, R.B. Olsen and O.V.
Pictet {\em An Introduction to High-Frequency Finance}, Academic
Press, New York, (2001).

\bibitem{Arch}
R.~F. Engle, {\em Econometrica} {\bf 50}, 987 (1982).

\bibitem{Garch}
T. Bollerslev, {\em J. Econometrics} {\bf 31}, 307 (1986).

\bibitem{Heston}
S.L. Heston, Rev. Financial Studies {\bf 6}, 327 (1993).

\bibitem{Hull-White}
 H. Hull and J. White {\em J. Finance} {\bf XLII} 281, (1987).

\bibitem{BouchaudCont}
J.-P. Bouchaud and R. Cont,
Eur. Phys. J. B {\bf 6}, 543 (1998).

\bibitem{OltreNES}
P. D. Ditlevsen, H. Svensmark and S. Johnsen, Nature \textbf{379},
810 (1996); P. D. Ditlevsen, Geophys. Res. Lett. \textbf{26}, 1441
(1999); P. D. Ditlevsen, M. S. Kristensen, and K. K. Andersen, J.
Climate \textbf{18}, 2594 (2005); Xie C.W., and Mei D. C., Chin.
Phys. Lett. \textbf{20} (6), 813 (2003);

\bibitem{BonannoPhysicaA}
S. Miccich\`e, G. Bonanno, F. Lillo and R.N. Mantegna,
Physica A {\bf 314}, 756, (2002).

\bibitem{Yakovenko}
A.A. Dragulescu and V.M. Yakovenko,
Quantitative Finance {\bf 2}, 443 (2002).

\bibitem{BonannoFNL}
G. Bonanno and B. Spagnolo, Fluctuation and Noise Letters {\bf 5},
L325, (2005); {\em Stochastic Models and Escape Times of Financial
Markets}, Mod. Probl. Stat. Phys. \textbf{4}, 122 (2005).

\bibitem{Silva}
A. Christian Silva, {\em Application of Physics to Finance and
Economics: Returns, Trading Activity and Income}, cond-mat/0507022
(2005).

\bibitem{BS}
F.Black and M.Scholes, Journal of Political Economy \textbf{81}, 637
(1973).

\bibitem{Merton}
R. C. Merton, Bell Journal of Economics and Management Science
\textbf{4}, 141 (1973).

\bibitem{Cox}
J.Cox, J.Ingersoll, and S.Ross, Econometrica \textbf{53}, 385
(1985).

\bibitem{Chalasani}
P. Chalasani and S. Jha, \emph{Steven Shreve: Stochastic Calculus
and Finance}, http://www.stat.berkeley.edu/users/evans/shreve.pdf.

\bibitem{Hanggi}
P. H$\ddot{a}$nggi, P. Talkner and M. Borkovec, Rev. Mod. Phys.
\textbf{62}, 251 (1990); E. Pollak and P. Talkner, Chaos
\textbf{15}, 026116 (2005).

\bibitem{Gardiner}
C.~W.~Gardiner {\em Handbook of stochastic methods for physics,
chemistry and the natural sciences}, {\em Springer, Berlin, 1993}.

\bibitem{NES03}
A. Fiasconaro, D. Valenti, B. Spagnolo, Physica A \textbf{325}, 136
(2003); A. Fiasconaro, B. Spagnolo, S. Boccaletti, Phys. Rev.
E\textbf{72}, 061110(5) (2005).

\bibitem{Fouque}
J.P. Fouque, G. Papanicolau and K.R. Sircar {\em Derivatives in
financial markets with stochastic volatility}, (Cambridge University
Press, Cambridge, 2000).

\bibitem{NumRec}
W. H. Press, S.A. Teukolsky, W. T. Vetterling, B.P. Flannery,
\emph{Numerical Recipes}, (Cambridge University Press, Cambridge,
1992).

\bibitem{Lil00}
F. Lillo and R. N. Mantegna, Eur. Phys. J. B \textbf{15}, 603
(2000).

\bibitem{LiMa00}
F. Lillo and R. N. Mantegna, Phys. Rev. E \textbf{62}, 6126
(2000).

\bibitem{Bon_PRE_03}
G. Bonanno, G. Caldarelli, F. Lillo, and Rosario N. Mantegna,  Phys.
Rev. E \textbf{68}, 046130 (2003).

\bibitem{Bon_EPJB04}
G. Bonanno, G. Caldarelli, F. Lillo, S. Miccich\`e, N. Vandewalle,
and R. N. Mantegna, Eur. Phys. J. B \textbf{38}, 363 (2004).

\end{thebibliography}
\end{document}